\documentclass[12pt]{iopart}

\usepackage{booktabs}
\usepackage{indentfirst}
\usepackage{diagbox}
\usepackage{pdflscape}

\expandafter\let\csname equation*\endcsname\relax
\expandafter\let\csname endequation*\endcsname\relax

\usepackage{graphicx}
\usepackage{siunitx}
\usepackage{tabularx}
\usepackage{enumitem}
\usepackage[font=small,labelfont=bf]{caption}
\usepackage{amssymb}

\usepackage[round]{natbib}
\usepackage{cite}

\usepackage[colorlinks=false]{hyperref}
\hypersetup{
  hidelinks,
  citecolor=black,
  linkcolor=black,
  urlcolor=black}

\usepackage[acronym,shortcuts]{glossaries}
\newacronym{FOV}{FOV}{field of view}
\newacronym{HMD}{HMD}{head-mounted display}
\newacronym{MAFC}{MAFC}{multiple-alternative forced-choice}
\newacronym{OLS}{OLS}{ordinary least squares}
\newacronym{RP}{RP}{retinitis pigmentosa}
\newacronym{SPV}{SPV}{simulated prosthetic vision}
\newacronym{VR}{VR}{virtual reality}

\usepackage{xcolor}

\begin{document}

\title[Simulated prosthetic vision confirms effective raster patterns]{Simulated prosthetic vision confirms checkerboard as an effective raster pattern for epiretinal implants} 

\author{Justin M. Kasowski\textsuperscript{1}, Apurv Varshney\textsuperscript{2}, Roksana Sadeghi\textsuperscript{2}, and Michael Beyeler\textsuperscript{2,3}}
\address{
    \textsuperscript{1}Interdepartmental Graduate Program in Dynamical Neuroscience, University of California, Santa Barbara, CA, 93106 \\
    \textsuperscript{2}Department of Computer Science, University of California, Santa Barbara, CA, 93106 \\
    \textsuperscript{3}Department of Psychological \& Brain Sciences, University of California, Santa Barbara, CA, 93106
}
\ead{\url{mbeyeler@ucsb.edu}}

\begin{abstract}
\emph{Objective:} Spatial scheduling of electrode activation (``rastering'') is essential for safely operating high-density retinal implants, yet its perceptual consequences remain poorly understood. This study systematically evaluates the impact of raster patterns, or spatial arrangements of sequential electrode activation, on performance and perceived difficulty in simulated prosthetic vision (SPV). By addressing this gap, we aimed to identify patterns that optimize functional vision in retinal implants.

\emph{Approach:} Sighted participants completed letter recognition and motion discrimination tasks under four raster patterns (horizontal, vertical, checkerboard, and random) using an immersive SPV system. The simulations emulated epiretinal implant perception and employed psychophysically validated models of electrode activation, phosphene appearance, nonlinear spatial summation, and temporal dynamics, ensuring realistic representation of prosthetic vision. Performance accuracy and self-reported difficulty were analyzed to assess the effects of raster patterning.

\emph{Main Results:} The checkerboard pattern consistently outperformed other raster patterns, yielding significantly higher accuracy and lower difficulty ratings across both tasks. The horizontal and vertical patterns introduced biases aligned with apparent motion artifacts, while the checkerboard minimized such effects. Random patterns resulted in the lowest performance, underscoring the importance of structured activation. Notably, checkerboard matched performance in the ``No Raster" condition, despite conforming to groupwise safety constraints.

\emph{Significance:} This is the first quantitative, task-based evaluation of raster patterns in SPV. Checkerboard-style scheduling enhances perceptual clarity without increasing computational load, offering a low-overhead, clinically relevant strategy for improving usability in next-generation retinal prostheses.
\end{abstract}

%
\vspace{2pc}

\newpage
\section{Introduction}

Retinal degenerative diseases, such as \ac{RP} and age-related macular degeneration, are major causes of severe vision loss, often diminishing quality of life and personal independence \citep{hamel_retinitis_2006, sainohira_quantitative_2018, prem_senthil_seeing_2017}. 
While early-stage \ac{RP} may benefit from emerging treatments like gene and stem cell therapies \citep{russell_efficacy_2017, da_cruz_phase_2018}, electronic visual prostheses remain a critical option for individuals in the advanced stages of degeneration \citep{beyeler_learning_2017}.

Visual prostheses function by capturing visual input, typically with an external camera, and converting it into electrical signals delivered to implanted microstimulators \citep{weiland_electrical_2016,fernandez_development_2018}. 
These devices stimulate surviving neurons in the retina or visual cortex, evoking the perception of flashes of light (\emph{phosphenes}). 
Among retinal devices, the Argus II Retinal Prosthesis System (Vivani Medical, Inc.; formerly Second Sight Medical Products, Inc.) was the first to achieve regulatory approval \citep{luo_argusr_2016} and has been implanted in 388 individuals worldwide (personal communication with Cortigent, Inc., 2024).
Currently nearing commercialization is PRIMA \citep{lorach_photovoltaic_2015,palanker_photovoltaic_2020}, a compact subretinal implant with a photovoltaic design that eliminates the need for external cables.
Originally commercialized by Pixium Vision, the technology has since been acquired by Science Corporation, which is now advancing its development.
Meanwhile, cortical prostheses with higher electrode counts are being developed to bypass the retina entirely, potentially providing a treatment option for a wide range of blindness causes \citep{chen_shape_2020, jung_stable_2024, musk_integrated_2019}.

Advancing high-resolution prostheses presents significant challenges. 
Safety guidelines established by regulatory agencies, such as those from the US Food and Drug Administration (FDA), limit the number of electrodes that can be activated simultaneously or in close succession to prevent adverse events, neural damage, and electrode degradation.
While these constraints are essential, they also hinder the resolution required for complex functional tasks.

One strategy to address these limitations involves preprocessing the visual input, using methods like edge detection or semantic segmentation to simplify scenes and emphasize task-relevant features \citep{parikh_saliency-based_2010, vergnieux_simplification_2017, sanchez-garcia_semantic_2020, han_deep_2021, rasla_relative_2022}. 
Although preprocessing may enhance perceptual clarity by emphasizing salient or task-relevant features of the scene \citep{beyeler_towards_2022}, safety constraints still necessitate activating only a subset of electrodes at any given moment.

Existing prostheses, such as Argus II, address this by activating subsets of electrodes (\emph{timing groups}; Second Sight Surgeon Manual, \citeyear{second_sight_argus_2013}) in rapid temporal succession. 
Inspired by raster scanning in display technologies, this approach sequentially segments the visual field into strips, aiming to create the perception of a coherent visual frame.
Raster patterns, which define the spatial arrangement and activation order of these electrode groups, are typically chosen randomly or adjusted heuristically based on limited user feedback \citep{second_sight_argus_2013}.
While these approaches address safety concerns, they lack systematic evaluation, leaving open questions about how specific patterns influence perceptual and behavioral outcomes.
This is particularly relevant as both retinal and cortical implants evolve to include hundreds or thousands of electrodes.

Direct evaluation of raster strategies in real implant users is currently infeasible: no commercial retinal prostheses are available, and clinical testing is limited by safety constraints, device variability, and small sample sizes.
\Ac{SPV} in immersive \ac{VR} offers a scalable, risk-free alternative, enabling controlled experiments with \emph{virtual patients} in realistic environments \citep{hayes_visually_2003, dagnelie_real_2007, kasowski_immersive_2022}. 
To ensure realism, simulations must be grounded in device-specific models. Here we focus on epiretinal implants because they are the best studied, with psychophysically validated models of phosphene appearance \citep{beyeler_model_2019, granley_computational_2021}, temporal dynamics \citep{hou_predicting_2024}, and nonlinear spatial summation \citep{hou_axonal_2024}. These models provide a principled foundation for evaluating stimulation strategies such as raster scheduling.

In this study, we systematically evaluated four raster patterns (horizontal, vertical, checkerboard, and random) using \ac{SPV} to assess their effects on letter recognition and motion discrimination tasks. 
We also evaluated the influence of raster frame rate and the number of electrodes per raster group in a follow-up experiment.
By analyzing task performance and perceptual biases across these configurations, we aimed to identify patterns that optimize usability while adhering to regulatory safety constraints.

Although checkerboard-like scheduling strategies are common in engineering and display systems, their impact on human perception in the context of visual prostheses has not been rigorously tested. Prior publications have mentioned raster scheduling only in passing, typically without quantitative evaluation or behavioral metrics. To our knowledge, no study has (i) systematically compared multiple spatial raster patterns, (ii) assessed behavioral consequences across multiple tasks, and (iii) incorporated realistic spatiotemporal phosphene dynamics into the simulation. Our work addresses this gap by providing the first data-driven rationale for selecting a checkerboard raster pattern over the linear or random patterns currently used in commercial devices.

Because this approach generalizes across tasks and requires no per-frame computation or patient-specific calibration, it offers a low-overhead, high-impact modification to existing stimulation protocols. As such, it represents a translationally relevant solution for improving usability in current and future implant systems.

As high-resolution prosthetic technologies for retinal and cortical applications continue to develop, refining electrode activation strategies will be essential to bridge the gap between engineering advancements and the practical needs of individuals with vision loss \citep{nadolskis_aligning_2024}.
Simulated experiments like this study represent an important step toward refining these systems and advancing their clinical viability.

\clearpage
\section{Methods}

\subsection{Participants}
\label{sec:participants}

We recruited 58 participants with normal or corrected-to-normal vision (ages 18–29; $M = 19.76$, $SD = 2.78$; 34 female, 24 male) from the Department of Psychological \& Brain Sciences research participant pool at the University of California, Santa Barbara (UCSB). 
Participants served as ``virtual patients'' \citep{kasowski_immersive_2022} in simulated prosthetic vision experiments.
Ten participants had never used \ac{VR} before.
To minimize the risk of cybersickness or discomfort, participants prone to light sensitivity or flashing lights were excluded. 

The study was approved by UCSB's Institutional Review Board. The research was conducted in accordance with the principles embodied in the Declaration of Helsinki and in compliance with local statutory requirements.

All participants provided written informed consent prior to participation. Participants were fully debriefed after completing the study.

\subsection{Simulated prosthetic vision (SPV)}
\label{sec:spv}

We used the open-source Unity toolbox \texttt{BionicVisionXR} \citep{kasowski_immersive_2022} to simulate prosthetic vision in an immersive \ac{VR} environment.
Participants viewed stimuli through an HTC VIVE Pro Eye head-mounted display, with phosphene appearance simulated using psychophysically validated models of epiretinal implants \citep{horsager_predicting_2009,beyeler_model_2019,granley_computational_2021}. This approach modeled spatiotemporal dynamics and gaze-contingent rendering (explained further below) to approximate the visual experiences of epiretinal prosthesis users. 
To simulate group-based stimulation, we used the updated ``axon map'' model, which captures nonlinear spatial interactions and more accurately predicts the shape distortions caused by simultaneous multi-electrode activation in epiretinal implants~\citep{hou_axonal_2024}.

To maintain generalizability while aligning with near-future epiretinal implants, we simulated a $10 \times 10$ epiretinal electrode array centered over the fovea, inspired by the Argus II implant \citep{luo_argusr_2016}. 
Electrode pitch was fixed at \SI{400}{\micro\metre}, similar to Argus II and supporting a field of view of approximately \SI{15}{\degree}.
Scene content dynamically updated based on participants' head and eye movements, providing a realistic approximation of user interaction with prosthetic vision.

\subsection{Raster patterns}
\label{sec:raster}

Visual prostheses must adhere to strict safety guidelines, including limits on electrode charge density and simultaneous current. 
These restrictions ensure safety but typically prevent all electrodes from being activated simultaneously.
To comply, implants such as the Argus II divide electrodes into ``timing groups'' (Second Sight Surgeon Manual, \citeyear{second_sight_argus_2013}), stimulated in rapid succession.
For this study, we use the term \emph{raster pattern} to refer to the spatial arrangement and activation sequence of these timing groups, analogous to raster line scanning in display systems.

\begin{figure}[!t]
\centering
    \includegraphics[width=\linewidth]{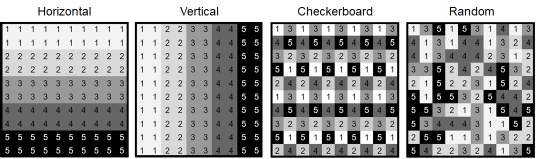}
    \caption{
        The four raster patterns tested in the study. 
        Electrodes were divided into five timing groups (labeled 1--5) that were sequentially activated every \SI{44.4}{\milli\second} (i.e., an effective per-group raster rate of \SI{22.5}{\hertz}), thereby completing a full raster cycle every \SI{222}{\milli\second} (\SI{4.5}{\hertz}).
        \emph{Horizontal:} Electrodes grouped into two adjacent rows, activated sequentially from top to bottom.
        \emph{Vertical:} Electrodes grouped into two adjacent columns, activated sequentially from left to right.
        \emph{Checkerboard:} Designed to maximize the distance between active electrodes.
        \emph{Random:} Electrode groupings were re-randomized every five frames, introducing maximal temporal variability with no spatial constraints. This served as a deliberately unstructured baseline to test the benefit of systematic spatial organization.
    }
    \label{fig:raster-patterns}
\end{figure}

To systematically evaluate performance under different raster patterns, we tested four configurations with varying spatial organization and randomness (Figure~\ref{fig:raster-patterns}).

\begin{itemize}
    \item \textbf{Horizontal:} Electrodes grouped in two adjacent rows, stimulated sequentially from top to bottom. This is believed to be equivalent to the default timing groups in Argus II (Second Sight Surgeon Manual, \citeyear{second_sight_argus_2013}). 
    \item \textbf{Vertical:} Electrodes grouped in two adjacent columns, stimulated from left to right.  This orthogonal arrangement allows for comparisons with the horizontal pattern and is designed to test whether raster directionality influences perceptual outcomes.
    \item \textbf{Checkerboard:} Electrodes were arranged to maximize spatial separation across successive activations, forming a checkerboard-like pattern. To transition between groups while avoiding apparent motion, the pattern was shifted using a non-linear strategy designed to preserve spatial separation. This approach minimized directional biases by avoiding simple linear shifts (e.g., one-pixel over) that could introduce apparent motion, instead ensuring each shift maintained balanced spacing and reduced perceptual interference between phosphenes.
    \item \textbf{Random:} Electrode activation randomized every five frames, with no fixed spatial structure to assess the importance of structured activation.
\end{itemize}

Each raster pattern divided the 100-electrode array into five spatially distinct groups, with 20\% of the electrodes activated per group. 
These groups were activated in rapid succession, completing a full raster cycle at \SI{4.5}{Hz}; that is, the array was refreshed every \SI{222}{ms}, with a \SI{44.4}{ms} delay between groups (i.e., an effective per-group rate of \SI{22.5}{\hertz}). This rate, while slower than the $2$-\SI{3}{ms} inter-group intervals typical of current devices, falls within inter-pulse delays reported in prior Argus II studies ($25$–$\SI{83}{ms}$) \citep{yucel_factors_2022}.

When applied to a visual stimulus, the raster pattern divides the scene into discrete regions rendered sequentially by the electrode groups (illustrated in Figure~\ref{fig:raster-patterns}). 
If the stimulation sequence is sufficiently rapid, temporal integration by the visual system may combine these fragments into a coherent percept \citep{rashbass_visibility_1970,efron_conservation_1973}. 
However, factors such as phosphene fading, persistence, and neural adaptation \citep{avraham_retinal_2021,hou_predicting_2024} complicate this process, and the influence of raster patterns on perceptual clarity and behavior remains unclear.

Supplemental Video~1 demonstrates a raster pattern with the temporal model enabled, while Supplemental Video~2 illustrates the same pattern without temporal dynamics. 
Notably, the temporal model mitigates abrupt transitions between raster groups, producing a smearing effect that enhances perceptual smoothness. 
This distinction is critical, as vertical and horizontal patterns without the temporal model can appear jarring and visually uncomfortable, highlighting the importance of incorporating temporal dynamics in simulated prosthetic vision studies.

\begin{figure}[!ht]
 \centering
    \includegraphics[width=\linewidth]{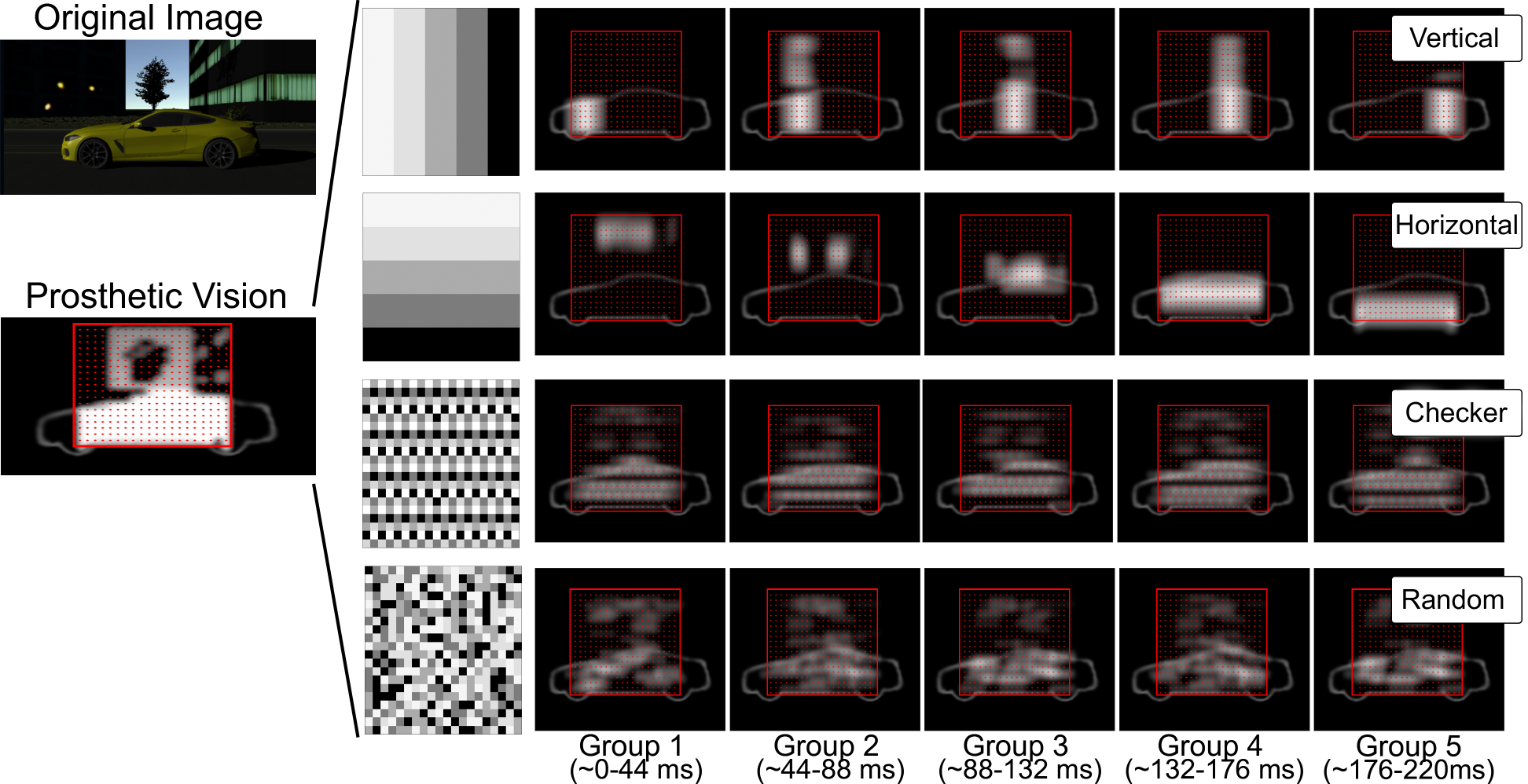}
    \caption{
        \textbf{Raster patterns in simulated prosthetic vision.}
        \emph{Left:} The original image is converted into an electrode activation pattern for simulating prosthetic vision. Red dots in the ``Prosthetic Vision'' overlay indicate electrode locations.
        \emph{Right:} To limit simulated current per frame, the 100 electrodes were divided into five raster groups, shown by the five gray levels in the leftmost panel. Four raster patterns were tested: vertical (groups scanned left to right), horizontal (top to bottom), checkerboard (maximally spaced electrodes per frame), and random (new group assignments every five frames). The full array refreshed at $\SI{4.5}{\hertz}$, meaning each raster group was activated once every \SI{222}{\milli\second}, resulting in an effective per-group rate of $\SI{22.5}{\hertz}$.
    }

    \label{fig:raster}
\end{figure}

\subsection{Tasks}
\label{sec:tasks}

Two eight-alternative forced-choice tasks were designed to evaluate the effects of different raster patterns on visual performance:
\begin{itemize}
    \item \textbf{Letter identification:} Modeled after \citet{cruz_argus_2013,kasowski_immersive_2022}, participants viewed one of eight Snellen optotypes (C, D, E, F, L, O, P, T) subtending \SI{41.1}{\degree} of visual angle. Each joystick direction corresponded to a letter (e.g., forward-left for ``C''), confirmed with a trigger button.
    \item \textbf{Motion discrimination:} Modeled after \citet{dorn_detection_2013}, participants viewed five-second videos generated using pulse2percept \citep{beyeler_pulse2percept_2017}, showing a single bar moving perpendicular to its orientation. Joystick directions indicated motion (e.g., forward-left for up/left), confirmed with a trigger button.
\end{itemize}

Both tasks were presented in a \ac{VR} environment, with stimuli displayed on a virtual monitor in a darkened room. 
Each stimulus was shown for \SI{5}{\second}, during which participants could use head and eye movements to scan the scene. 
Participants selected their response using a joystick, with each of the eight directions corresponding to a specific choice. 
If no response was made during the stimulus presentation, the scene went black, and participants were required to make a decision. 
Immediate feedback (correct/incorrect) was provided after each trial, and participants had up to \SI{5}{\second} to finalize their response.

\begin{figure}[!t]
    \centering
    \includegraphics[width=\linewidth]{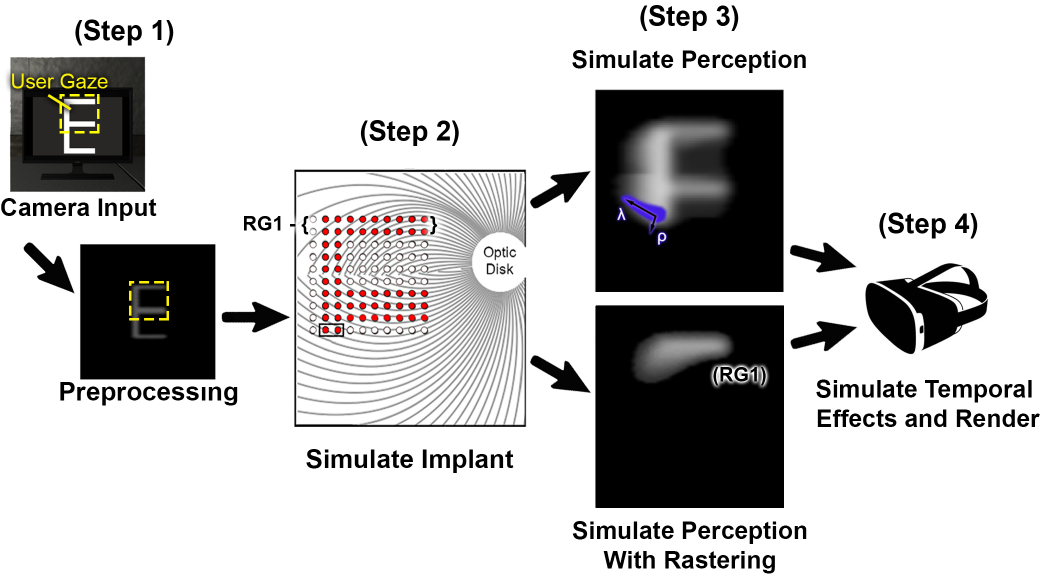}
    \caption{
        \textbf{Simplified overview of the simulated prosthetic vision model.} 
        \emph{Step~1:} Unity's virtual camera captures the scene and preprocesses it (grayscale, Gaussian subsampling).
        \emph{Step~2:} Electrode activation is determined based on the visual input as well as the placement of the simulated retinal implant. In the current study, a $3 \times 3$ Gaussian blur was applied to the preprocessed image to average the grayscale values around each electrode's location in the visual field. This gray level was then interpreted as a current amplitude delivered to a particular electrode in the array. Electrodes are stimulated in close temporal succession according to their raster group;         
        Raster Group 1 (RG1) electrodes are highlighted.
        \emph{Step~3:} Perception is simulated either with (\emph{bottom}) or without (\emph{top}) rastering. Phosphene shape is determined by parameters $\rho$ (spread) and $\lambda$ (elongation). The highlighted shape represents the percept generated by two active electrodes in the horizontal raster condition, where only 20 out of 100 electrodes are active per frame.
        \emph{Step~4:} Temporal dynamics are applied, and the final percept is rendered to the headset.}
    \label{fig:overview}
\end{figure}

\subsection{Simulation pipeline}

The raster patterns were overlaid onto scenes, and visual input was processed through the following steps (Fig.~\ref{fig:overview}):

\begin{enumerate}[itemsep=0.25em,label=\arabic*.]
    \item \textbf{Image acquisition:} Unity’s virtual camera captured a \SI{60}{\degree} field of view, rendered at \SI{90}{Hz}.
    \item \textbf{Image processing:} Frames were downscaled to $200 \times 200$ pixels, converted to grayscale, and smoothed with a $3 \times 3$ Gaussian kernel.
    \item \textbf{Electrode activation:} Pixel intensities nearest to each electrode were used to compute activation levels. Only electrodes in the current raster group were stimulated (explained in Section~\ref{sec:raster}).
    \item \textbf{Spatial effects:} Phosphene shapes were modeled using the axon map \citep{beyeler_model_2019,granley_computational_2021}, simulating elongated phosphenes aligned with retinal ganglion cell axons (Section~\ref{sec:spatial-model}).
    \item \textbf{Temporal effects:} A temporal model \citep{horsager_predicting_2009} simulated phosphene fading and persistence by accounting for charge accumulation and decay (Section~\ref{sec:temporal-model}).
    \item \textbf{Gaze-contingent rendering:} The implant location dynamically shifted based on gaze position, ensuring the scene remained aligned with participants’ fixation (Section~\ref{sec:gaze}).
\end{enumerate}

This pipeline integrated spatial and temporal distortions to provide a realistic approximation of prosthetic vision \citep{kasowski_immersive_2022}.

\subsubsection{Spatial distortions}
\label{sec:spatial-model}

The shape of phosphenes in epiretinal devices is influenced by the retinal ganglion cell axons, which traverse the retina in curved paths \citep{rizzo_perceptual_2003,beyeler_model_2019}. 
We used the axon map model to simulate these distortions \citep{beyeler_model_2019,granley_computational_2021}.
Each electrode activated a region of the retina defined by Gaussian falloff parameters $\rho$ (spread) and $\lambda$ (elongation).
The instantaneous brightness $b_I$ of each pixel $(r,\theta)$ in the percept was computed according to:
\begin{equation}
    b_I = \max_{p \in R(\theta)} \sum_{e \in E} \exp \bigg( \frac{-d_e^2}{2\rho^2} + \frac{-d_\mathrm{soma}^2}{2\lambda^2} \bigg),
    \label{eq:axon-map}
\end{equation}
where $R(\theta)$ is the path of the axon terminating at retinal location $(r,\theta)$, $p$ is a point along the path, $d_e$ is the distance from $p$ to the stimulating electrode $e$, and $d_\mathrm{soma}$ is the distance along the axon from $p$ to the cell body. 

This formulation introduces nonlinear spatial summation by computing the brightness at each location as the maximum over the axon path rather than a linear sum. This mechanism, supported by recent psychophysical studies \citep{hou_axonal_2024}, captures perceptual interactions between electrodes under simultaneous stimulation, including merging, elongation, and asymmetric brightness distributions.

Values for $\rho$ and $\lambda$ were calibrated for each participant using a staircase procedure (Section~\ref{sec:procedure}).

\subsubsection{Temporal distortions}
\label{sec:temporal-model}

To model temporal dynamics, we used a simplified variant of the \citet{horsager_predicting_2009} model, which incorporates two coupled leaky integrators to simulate neural desensitization $n(t)$ and phosphene brightness $b(t)$. The governing equations were:
\begin{align}
    \frac{dn(t)}{dt} &= -\tau_n n(t) + b_I(t), \label{eq:charge-accumulation}\\
    \frac{db(t)}{dt} &= -\tau_b b(t) - \alpha n(t) + b_I(t), \label{eq:phosphene-brightness}
\end{align}
where $b_I(t)$ was the instantaneous brightness (from the spatial model) calculated at time $t$.
Parameter values (for \SI{5}{\hertz}: $\tau_n = \SI{0.2}{\second}$, $\tau_b = \SI{5}{\second}$, and $\alpha = 0.2$; for 20Hz: $\tau_n = \SI{0.2}{\second}$, $\tau_b = \SI{5}{\second}$, and $\alpha = 0.25$) were fitted to reproduce temporal fading and persistence effects reported by Subject 5 of \citet{perez_fornos_temporal_2012} (see their (Figures 3 and 4).

\subsubsection{Gaze-contingent phosphene rendering}
\label{sec:gaze}

All raster patterns in this study were rendered gaze-contingently to simulate a more realistic and adaptive prosthetic vision experience. Participants' eye movements were recorded in real time using the HTC Vive Pro Eye. For each video frame, the input image was shifted to align the participant’s current fixation point with the center of the simulated implant. This ensured that stimulation patterns remained stable in retinal coordinates, regardless of eye movements.

Rendering in retinal coordinates is essential for biologically plausible simulation: it reflects how visual input interacts with localized neural adaptation and temporal filtering in the retina. By contrast, screen-fixed stimulation would result in perceptual smearing or spatial distortions during eye movements, which is an unrealistic assumption for modern prosthetic systems.

The importance of gaze-contingent stimulation is increasingly recognized in the field~\citep{caspi_eye_2018,caspi_eye_2021,paraskevoudi_eye_2019}, and future implant systems are expected to incorporate this feature, either via on-chip photodiodes or through external eye-tracking hardware.

Eye-tracking precision in our setup was approximately \SI{1.9}{\degree} on average, with \SI{94}{\percent} of samples falling within \SI{5}{\degree} of the ground-truth target during both fixation and pursuit tasks (see Supplementary Materials).

\subsection{Procedure}
\label{sec:procedure}

Participants completed demographic and screening surveys prior to attending the session, which assessed eligibility and collected background information.
Upon arrival, they donned the headset and completed HTC's eye calibration procedure. 
For a subset of participants ($n = 30$), additional precision checks were conducted using a ``follow the dot" task (see Supplementary Materials). 
Participants unfamiliar with \ac{VR} were provided extra time to acclimate using a virtual test room and joystick controls.

\subsubsection{Training phase}
To help participants adapt to the spatial distortions introduced by simulated prosthetic vision, a structured training phase was implemented. 
Training consisted of three sets of five trials with incrementally increasing levels of distortion.
Initial trials featured low distortion parameters ($\lambda = 50$, $\rho = 150$) and a pixel-like image rendered using 400 ($20 \times 20$) electrodes spaced \SI{300}{\micro\meter} apart. 
This provided a large field of view and familiarized participants with the interface. 
Subsequent trials used a reduced electrode count (100 electrodes in a $10 \times 10$ grid) spaced \SI{400}{\micro\meter} apart and progressively increased the radial distortion ($\rho = 300$). 
By the final training level, distortion parameters ($\lambda = 1000$, $\rho = 300$) aligned with realistic values reported for current epiretinal devices \citep{beyeler_model_2019}, simulating a restricted $14.6^\circ \times 14.6^\circ$ field of view. 
Temporal effects, including rasterization, were disabled during training, but gaze-contingent rendering was active to simulate a dynamic viewing experience.

Training began with non-SPV trials (normal vision) before progressing to SPV conditions. 
Extensive piloting ensured that participants reached 70--90\% accuracy in the baseline (non-rastered) condition after training, providing a foundation for consistent performance during the experimental phase.

\subsubsection{Experimental phase}
Participants were randomly assigned to start with either the letter identification task or the motion discrimination task. 
All participants first completed a baseline condition (``No Raster''), where all electrodes were active on every frame but temporal effects were enabled (see Section~\ref{sec:temporal-model}). 
They then proceeded through the four raster patterns (horizontal, vertical, checkerboard, and random) in a counterbalanced order. 
Each condition comprised 48 trials, divided into six blocks of eight randomized stimuli. 
Blocks were presented without explicit markers, creating the impression of continuous randomization.

To assess whether raster rate meaningfully impacted performance, a follow-up experiment was conducted with a subset of participants ($n=10$) who repeated the letter task under three conditions: five groups at \SI{5}{Hz}, five groups at \SI{18}{Hz} (driven by the headset's \SI{11}{ms} refresh timing), and four groups at \SI{5}{Hz}. Only the ``Checkerboard'' and ``Random'' raster patterns were tested, and condition order was randomized.

Participants were allowed to abort the session at any time if they felt discomfort or cybersickness. 
Breaks were offered between blocks, and no participants required intervention. 
Each session lasted approximately two hours, including training, experimental trials, and debriefing.

\subsection{Data collection \& analysis}
\label{sec:data-analysis}

For each trial, we recorded task accuracy (correct/incorrect), self-reported difficulty ratings, and eye/head tracking data.

To assess overall learning effects, we fit a generalized linear mixed-effects model (GLMM) with a binomial distribution and logit link, using the \texttt{glmmTMB} package. The model included fixed effects for \texttt{Task}, \texttt{RasterSetting}, \texttt{Block}, and all interactions. A random intercept for each participant was included to account for individual variability:

\begin{equation*}
\texttt{correct} \sim \texttt{Task} \times \texttt{RasterSetting} \times \texttt{Block} + (1 \mid \texttt{Participant}).
\end{equation*}

To focus on asymptotic performance, we fit a second GLMM to Block 6 trials only (i.e., after learning plateaued), with fixed effects for \texttt{Task} and \texttt{RasterSetting}, and a random intercept for \texttt{Participant}:

\begin{equation*}
\texttt{correct} \sim \texttt{Task} \times \texttt{RasterSetting} + (1 \mid \texttt{Participant}).
\end{equation*}

For the follow-up experiment assessing the effects of raster frequency and group size, we used a third binomial GLMM with fixed effects for \texttt{RasterFrequency} (\SI{5}{\hertz} vs. \SI{18}{\hertz}), \texttt{RasterGrouping} (4 vs. 5 groups), and \texttt{RasterStrategy} (``Checkerboard'' vs. ``Random''), including all interactions:

\begin{equation*}
\texttt{correct} \sim \texttt{RasterFrequency} \times \texttt{RasterGroup} \times \texttt{RasterSetting} + (1 \mid \texttt{Participant}).
\end{equation*}

Post-hoc comparisons were performed using estimated marginal means with Tukey correction (\texttt{emmeans} package). Omnibus significance was assessed via Type III Wald chi-squared tests.

All analyses were conducted in R (v4.3.2).

\clearpage
\section{Results}

\begin{figure}[!b]
    \centering
    \includegraphics[width=\linewidth]{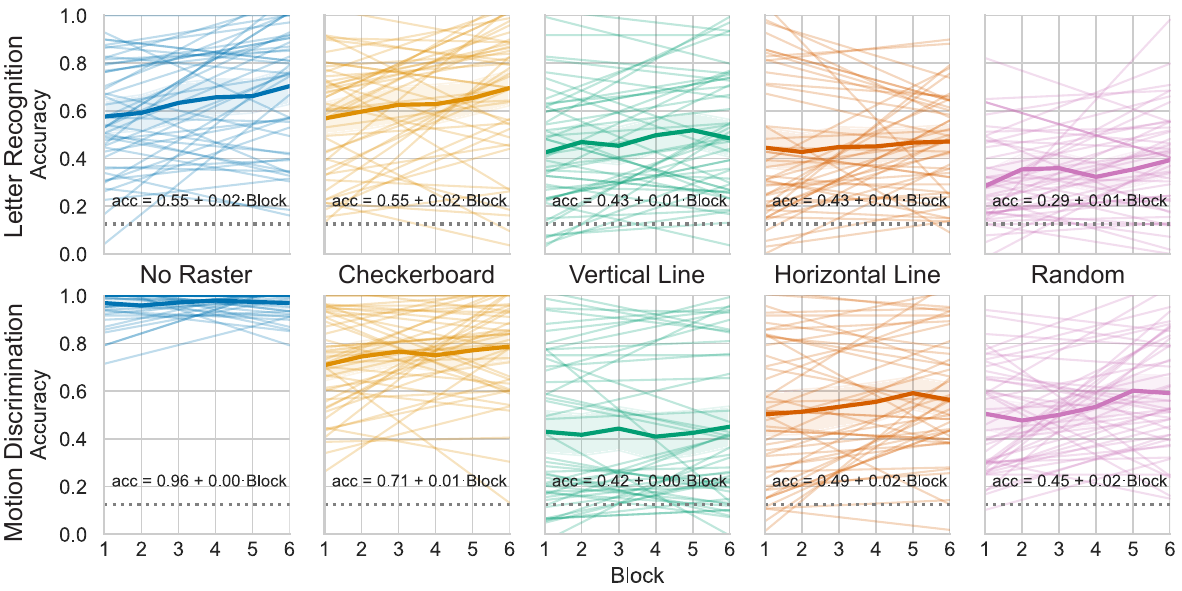}
    \caption{
        \textbf{Accuracy improves across blocks, with consistent group trends and variable individual learning.}
        Accuracy is plotted over six experimental blocks for each \texttt{RasterSetting} (columns) and Task (rows). 
        The ``No Raster'' condition served as a baseline and reached ceiling in the motion discrimination task.
        Thin colored lines represent linear fits for individual participants; thicker lines with shaded bands denote the group mean and 95\% confidence interval.
        The dotted line marks chance performance (1/8).
    }
    \label{fig:results-learning}
\end{figure}

\subsection{Performance improves across blocks}

To assess perceptual learning, we fit a generalized linear mixed-effects model (GLMM) to trial-level accuracy with a logit link, including fixed effects for \texttt{Task}, \texttt{RasterSetting}, \texttt{Block}, and all interactions, plus a random intercept for \texttt{Participant}.
The model converged successfully (AIC = 26241.3) and revealed a significant \texttt{Task} × \texttt{RasterSetting} interaction (e.g., ``TaskMotion'' × ``No Raster'': $\beta = 2.51$, $SE = 0.33$, $z = 7.65$, $p < .001$) and a positive main effect of \texttt{Block} ($\beta = 0.11$, $SE = 0.027$, $z = 4.12$, $p < .001$).
Because of the interaction we do not interpret the main effects of \texttt{Task} or \texttt{RasterSetting}.

Accuracy improved significantly across blocks, indicating robust perceptual learning for both tasks and all raster patterns (Figure~\ref{fig:results-learning}). In the letter recognition task, the checkerboard pattern showed both the highest initial accuracy and one of the steepest learning slopes, suggesting an immediate and compounding advantage. In the motion discrimination task, the ``No Raster'' condition reached ceiling early, while checkerboard again led among structured rasters.

Despite these consistent group-level trends, individual trajectories varied markedly. The random-intercept standard deviation ($\sigma = 0.66$ log-odds) implied a nearly twofold difference in baseline performance across participants (median odds-ratio $\approx 1.9$). While 83\% of subject-specific trends showed improvement over time, a minority exhibited stagnation or decline, which likely reflects differences in strategy, engagement, or calibration artifacts.

\subsection{Performance at asymptote reflects raster pattern quality}

To assess steady-state performance, we fit a logistic mixed-effects model to Block 6 accuracy across all participants (4,608 trials per raster), including fixed effects for \texttt{Task} and \texttt{RasterSetting} and their interaction, with a random intercept for \texttt{Participant}. 
The model converged successfully and showed a robust \texttt{Task} × \texttt{RasterSetting} interaction ($\chi^2(4) = 77.90$, $p < .001$).
Accordingly, we focus on Tukey-adjusted contrasts within each task and do not interpret the isolated main effects.

\begin{figure}[!tb]
    \centering
    \includegraphics[width=\linewidth]{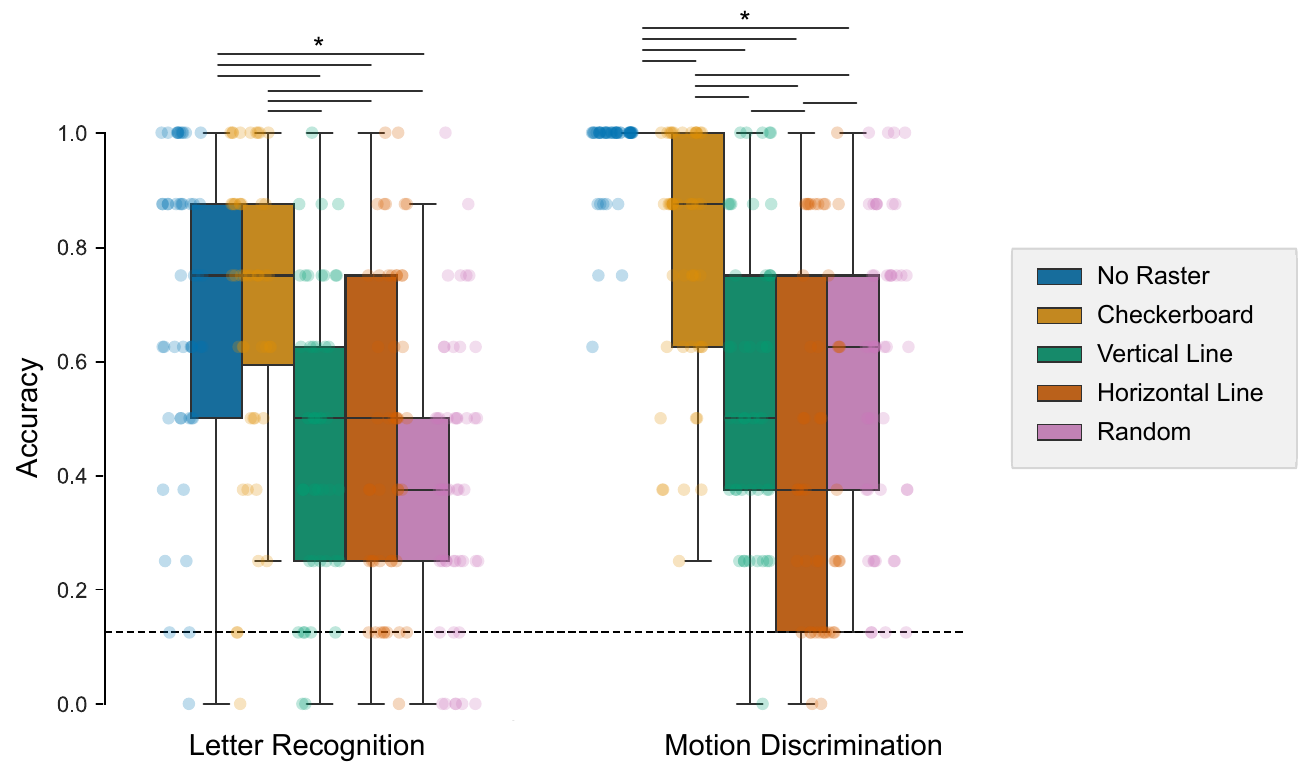}
    \caption{
        \textbf{Final block accuracy by raster pattern and task.} 
        Accuracy in Block~6 is shown for both tasks, grouped by RasterSetting. 
        Boxes show median and interquartile range; colored points represent individual participants.
        Horizontal bars reflect significant pairwise comparisons based on Tukey-adjusted contrasts from a binomial GLMM (* denotes $p<.05$).
        The dotted line marks chance performance (1/8).
    }
    \label{fig:results}
\end{figure}

In the letter recognition task, the checkerboard raster significantly outperformed all structured alternatives. Compared to horizontal lines, accuracy increased 3× ($\mathrm{OR}=3.00$, $z=5.93$, $p < .001$); versus vertical, 2.8× ($\mathrm{OR}=2.83$, $z=5.07$, $p < .001$); and versus random, 4.4× ($\mathrm{OR}=4.41$, $z=8.14$, $p < .001$). Critically, checkerboard matched the No Raster baseline ($\mathrm{OR}=0.86$, $z=-0.72$, $p=.95$), showing that spatial scheduling can preserve performance even under safety constraints.

In the motion discrimination task, the ``No Raster'' condition yielded ceiling-level accuracy and outperformed all other patterns. Still, checkerboard led among structured rasters, outperforming horizontal ($\mathrm{OR}=3.42$, $z=6.23$, $p < .001$), vertical ($\mathrm{OR}=5.76$, $z=8.10$, $p < .001$), and random ($\mathrm{OR}=2.93$, $z=5.55$, $p < .001$).

These results demonstrate that checkerboard rastering is the most effective of the structured options tested, offering a practical and perceptually robust alternative to the unstructured baseline.

\subsection{Systematic biases depend on stimulus and raster pattern}

\begin{figure}[!tp]
    \centering
    \includegraphics[width=\linewidth]{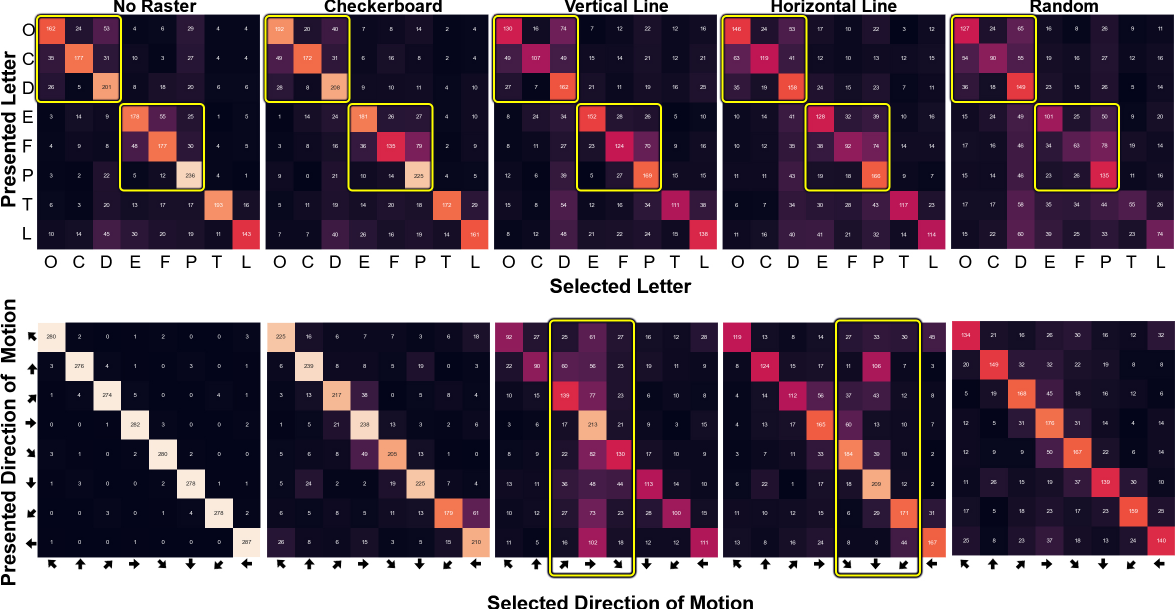}
    \caption{
        \textbf{Response bias by stimulus and raster pattern.} 
        The letter recognition task (top) exhibited grouping biases for visually similar letters ([C, D, O] and [E, F, P]). 
        In the motion discrimination task (bottom), apparent motion biases emerged for vertical and horizontal raster patterns, favoring the direction of electrode activation.
        Yellow rectangles highlight expected biases.
    }
    \label{fig:results-stimResponse} 
\end{figure}

To better understand the mechanisms underlying performance differences, we examined participants' response biases across raster patterns and tasks (Figure~\ref{fig:results-stimResponse}). Distinct and interpretable patterns of confusion emerged in both tasks.

In the letter task, participants were more likely to confuse visually similar letters regardless of raster pattern. However, confusion rates were lowest in the checkerboard and ``No Raster'' conditions, suggesting that structured rasters (especially horizontal and vertical) introduced additional interference.

In the motion task, horizontal and vertical rasters introduced systematic biases aligned with their activation direction. For example, the vertical raster led to more errors for rightward motion, consistent with downward apparent motion introduced by sequential column-wise stimulation. These biases were absent in the checkerboard and random conditions, indicating that spatial separation (checkerboard) and temporal randomness (random) helped preserve perceptual neutrality.

Interestingly, across both tasks, letters D and P were disproportionately chosen during incorrect responses. This likely reflects an ergonomic artifact: these letters were mapped to forward/backward joystick motions, which may have been more intuitive under uncertainty.

Together, these patterns support the idea that spatial layout influences not only performance but also perceptual interpretation. The checkerboard pattern minimized biases and helped preserve veridical perception across tasks.

\subsection{Self-reported difficulty ratings mirror performance trends}

\begin{figure}[tp!]
    \centering
    \includegraphics[width=\linewidth]{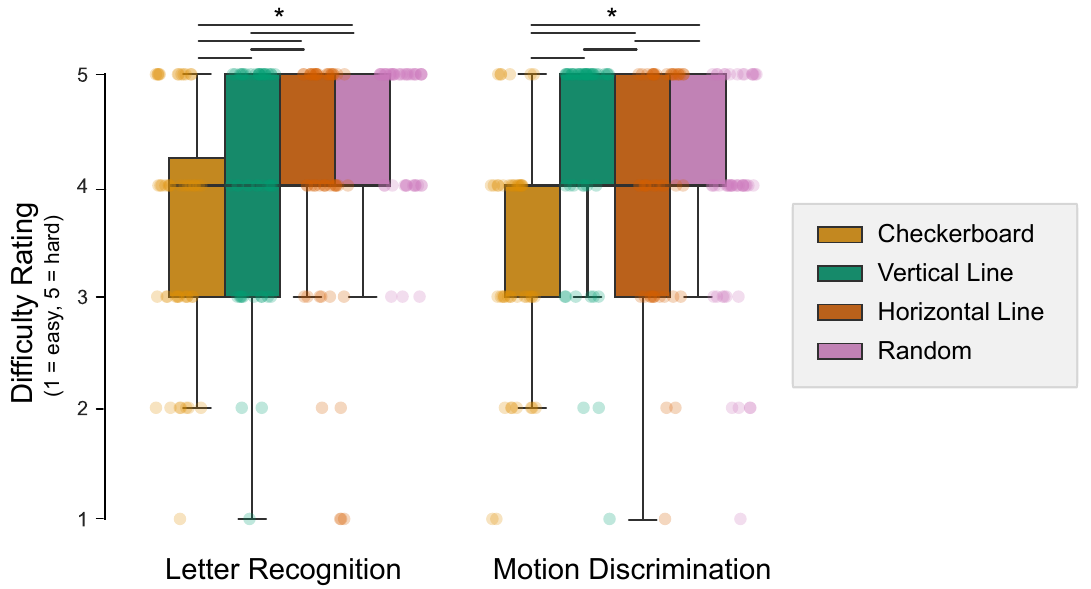}
    \caption{
        \textbf{Self-reported difficulty ratings by raster pattern and task.}
        Participants rated task difficulty on a 5-point scale (1 = easy, 5 = hard) after completing Block~6.
        Boxes indicate median and interquartile range; colored points show individual participants.
        Horizontal bars denote significant pairwise differences based on Tukey-adjusted contrasts (* denotes $p<.05$).
    }
    \label{fig:results-diff}
\end{figure}

Participants provided difficulty ratings after each block (1=easy, 5=difficult), allowing us to quantify subjective usability alongside objective performance (Figure~\ref{fig:results-diff}). Across both tasks, difficulty ratings mirrored accuracy: conditions that yielded higher performance were generally rated as easier.

In the letter task, checkerboard and ``No Raster'' conditions were rated as easiest, significantly outperforming horizontal ($p < .001$), vertical ($p < .001$), and random rasters ($p < .001$). In the motion task, ``No Raster'' again received the lowest difficulty ratings, followed by checkerboard ($p < .001$ vs. all other structured rasters).

These results reinforce the checkerboard pattern's practical utility: not only does it improve accuracy and reduce bias, it also reduces cognitive load as reported by users.

\subsection{Checkerboard performance is robust to frame rate and group size}

To evaluate the generality of our findings, a follow-up experiment tested whether raster frequency (\SI{5}{\hertz} vs.~\SI{18}{\hertz}) and number of groups (4 vs.~5) modulated performance. Ten participants repeated the letter task using the ``Checkerboard'' and ``Random'' raster patterns under these conditions.

A binomial GLMM revealed a significant interaction between raster frequency and raster type ($\hat{\beta} = -1.72$, $p < .001$); neither raster frequency nor group size showed a main effect.
Post-hoc Tukey contrasts confirmed that raising the array rate from \SI{5}{\hertz} to \SI{18}{\hertz} markedly improved the \emph{Random} raster ($\hat{\beta} = 2.46$, $SE = 0.45$, $z = 5.45$, $p < .0001$) but had no effect on \emph{Checkerboard} ($\hat{\beta} = 0.73$, $SE = 0.59$, $z = 1.26$, $p = .21$).  
Critically, even at \SI{18}{\hertz} the \emph{Random} raster remained inferior to \emph{Checkerboard}: log-odds difference of $1.05 \pm 0.52$ ($z = 2.03$, $p < .05$), corresponding to an odds-ratio of $\sim 2.9$.  Thus, higher refresh narrows but does not eliminate the checkerboard advantage.

Performance was likewise unchanged when the number of raster groups was reduced from five to four. 
Taken together, these results indicate that the checkerboard advantage derives primarily from its spatial layout rather than from temporal parameters such as update rate or group size, underscoring its potential suitability for a wide range of implant configurations.

\section{Discussion}

We systematically evaluated four raster patterns in a simulated prosthetic vision environment and found that the checkerboard layout consistently outperformed horizontal, vertical, and random strategies. It yielded higher accuracy and lower difficulty ratings across both letter recognition and motion discrimination tasks. By maximizing spatial separation between simultaneously activated electrodes, the checkerboard pattern reduced phosphene interference and preserved perceptual clarity, achieving performance levels comparable to the idealized ``No Raster'' condition, while still respecting regulatory safety constraints.

These findings offer immediate translational value. Unlike approaches that require individualized calibration or complex preprocessing, checkerboard-style rastering is hardware-agnostic, generalizes across tasks, and can be implemented with minimal overhead. As such, it represents a simple yet impactful optimization for current and future high-density visual prostheses.

\subsection{Checkerboard rastering improves accuracy and reduces perceptual artifacts}

Task performance depended critically on the spatial layout of electrode activation. The checkerboard pattern yielded the highest accuracy across both letter recognition and motion discrimination tasks (Figure~\ref{fig:results}), supporting the hypothesis that maximizing spatial separation between active electrodes reduces perceptual interference. This aligns with prior findings that closely spaced phosphenes tend to merge, degrading shape recognition \citep{horsager_spatiotemporal_2010, wilke_electric_2011, yucel_factors_2022, hou_axonal_2024}.

In the motion task, checkerboard rastering further reduced directional biases introduced by sequential activation in vertical and horizontal patterns. Its spatial balance mitigated apparent motion artifacts that can confound perceptual judgments.

These benefits are not tied to a specific device configuration: although our simulations used a \SI{400}{\micro\metre} pitch to match the Argus~II array, it is the relative spacing (rather than the absolute pitch) that determines raster effectiveness. As electrode density scales, group separation scales proportionally, making the checkerboard advantage generalizable across implant designs.

Participants' self-reported difficulty ratings further underscored the checkerboard pattern's advantages (Figure~\ref{fig:results-diff}). 
Lower difficulty ratings in both tasks suggest that this pattern not only enhances performance but also improves the subjective user experience, which is essential for promoting device usability and adoption \citep{beyeler_towards_2022, kasowski_systematic_2023, nadolskis_aligning_2024}.

\subsection{Limitations}

While this study highlights the advantages of the checkerboard raster pattern, several limitations must be acknowledged.

First, although devices like the Argus~II operate at millisecond-level temporal resolution, achieving such rapid rastering was not feasible with our VR-based setup. 
More importantly, this constraint mirrors an emerging challenge for future high-electrode-count implants: as the number of channels increases, it may become impractical to cycle through all electrode groups within a few milliseconds. This would effectively push raster update rates below the flicker fusion threshold, making the spatial arrangement of stimulation groups (and not just timing) a dominant factor in perception. Our findings are therefore most relevant to next-generation prosthetic systems, where optimizing spatial sampling strategies will be key to improving usability and performance.

To assess this directly, we conducted a follow-up experiment comparing \SI{5}{Hz} and \SI{18}{Hz} stimulation using both checkerboard and random raster patterns. Performance improved with higher frequency in the random condition but remained unchanged for checkerboard, suggesting that spatial layout (not frame rate) was the dominant factor in shaping perception. This robustness strengthens the translational relevance of checkerboard scheduling, particularly for high-resolution implants where rapid cycling may be infeasible.

Second, the study relied on simulated prosthetic vision with sighted participants. 
While our phosphene model is grounded in clinical data from Argus~II users~\citep{beyeler_model_2019, granley_computational_2021, hou_predicting_2024}, it also captures nonlinear spatial interactions that arise under simultaneous stimulation. Recent work shows that multi-electrode phosphenes are not linearly additive, and our ``axon map'' model provides a significantly better fit to user-drawn percepts than additive alternatives~\citep{hou_axonal_2024}. However, as with any study involving sighted participants, our simulations cannot reproduce the long-term cortical plasticity, perceptual reweighting, or compensatory strategies observed in blind prosthesis users~\citep{beyeler_learning_2017, mowad_compensatory_2020, castaldi_visual_2016, perez_fornos_temporal_2012, sadato_activation_1996}.
Validating these findings with implant recipients remains an essential next step.

Finally, the chosen tasks (letter recognition and motion discrimination) capture specific aspects of functional vision but do not encompass the full range of challenges faced by prosthesis users. 
Including tasks like navigation, object recognition, and dynamic scene analysis \citep{geruschat_flora_2015} could provide a more comprehensive evaluation of raster patterns.

\subsection{Future directions}

The findings from this study provide a foundation for optimizing raster patterns in visual prostheses and suggest promising directions for future research.

The checkerboard raster pattern consistently demonstrated superior performance across tasks, reducing perceptual interference and avoiding the apparent motion biases seen with vertical and horizontal patterns. 
This result underscores its potential as a design principle for next-generation prosthetic devices, particularly those with high electrode counts. 
Future work should investigate how the checkerboard pattern performs under more realistic temporal resolutions and with advanced device features, such as gaze-contingent updates and dynamic stimulation strategies.

Beyond validating these results with prosthesis users, expanding to more complex and ecologically valid tasks will help bridge the gap between controlled simulations and real-world usability. 
Investigating the interplay between raster patterns, electrode density, and field of view could reveal additional insights into optimizing stimulation strategies for practical use.

Finally, this study highlighted substantial individual variability in performance and learning rates. 
While our experimental design held the front-end image processing fixed to isolate the effects of raster scheduling, future work should explore how different encoding strategies (such as contrast normalization, semantic filtering, or task-adaptive preprocessing) interact with raster patterning. Our \ac{SPV} framework supports this kind of modular testing and could help identify display strategies that generalize more effectively across users and tasks \citep{beyeler_towards_2022, rasla_relative_2022, han_deep_2021}.

\section*{Acknowledgments}

We thank Jiaxin Su for coordinating participant scheduling during the follow-up experiment.
This work was supported by the National Eye Institute of the National Institutes of Health under Award Number R00-EY029329.
The content is solely the responsibility of the authors and does not necessarily represent the official views of the National Institutes of Health.

\section*{Data Availability}

The raw data for this study is publicly available on the Open Science Framework at \url{https://osf.io/nm6aw/}.
Prosthetic vision simulations were generated with \texttt{BionicVisionXR} \citep{kasowski_immersive_2022}, an open-source Unity toolbox available at \url{https://github.com/bionicvisionlab/BionicVisionXR}.

\pagebreak
\bibliographystyle{apalike}
\bibliography{2025-JNE-Raster}

\begin{thebibliography}{}

\bibitem[Avraham et~al., 2021]{avraham_retinal_2021}
Avraham, D., Jung, J., Yitzhaky, Y., and Peli, E. (2021).
\newblock Retinal prosthetic vision simulation: temporal aspects.
\newblock {\em Journal of Neural Engineering}.

\bibitem[Beyeler et~al., 2017a]{beyeler_pulse2percept_2017}
Beyeler, M., Boynton, G., Fine, I., and Rokem, A. (2017a).
\newblock pulse2percept: {A} {Python}-based simulation framework for bionic
  vision.
\newblock In {\em Proceedings of the 16th {Python} in {Science} {Conference}},
  pages 81--88, Austin, Texas. SciPy.

\bibitem[Beyeler et~al., 2019]{beyeler_model_2019}
Beyeler, M., Nanduri, D., Weiland, J.~D., Rokem, A., Boynton, G.~M., and Fine,
  I. (2019).
\newblock A model of ganglion axon pathways accounts for percepts elicited by
  retinal implants.
\newblock {\em Scientific Reports}, 9(1):1--16.

\bibitem[Beyeler et~al., 2017b]{beyeler_learning_2017}
Beyeler, M., Rokem, A., Boynton, G.~M., and Fine, I. (2017b).
\newblock Learning to see again: biological constraints on cortical plasticity
  and the implications for sight restoration technologies.
\newblock {\em Journal of Neural Engineering}, 14(5):051003.

\bibitem[Beyeler and Sanchez-Garcia, 2022]{beyeler_towards_2022}
Beyeler, M. and Sanchez-Garcia, M. (2022).
\newblock Towards a {Smart} {Bionic} {Eye}: {AI}-powered artificial vision for
  the treatment of incurable blindness.
\newblock {\em Journal of Neural Engineering}, 19(6):063001.
\newblock Publisher: IOP Publishing.

\bibitem[Caspi et~al., 2021]{caspi_eye_2021}
Caspi, A., Barry, M.~P., Patel, U.~K., Salas, M.~A., Dorn, J.~D., Roy, A.,
  Niketeghad, S., Greenberg, R.~J., and Pouratian, N. (2021).
\newblock Eye movements and the perceived location of phosphenes generated by
  intracranial primary visual cortex stimulation in the blind.
\newblock {\em Brain Stimulation}, 14(4):851--860.

\bibitem[Caspi et~al., 2018]{caspi_eye_2018}
Caspi, A., Roy, A., Wuyyuru, V., Rosendall, P.~E., Harper, J.~W., Katyal,
  K.~D., Barry, M.~P., Dagnelie, G., and Greenberg, R.~J. (2018).
\newblock Eye {Movement} {Control} in the {Argus} {II} {Retinal}-{Prosthesis}
  {Enables} {Reduced} {Head} {Movement} and {Better} {Localization}
  {Precision}.
\newblock {\em Investigative Ophthalmology \& Visual Science}, 59(2):792--802.

\bibitem[Castaldi et~al., 2016]{castaldi_visual_2016}
Castaldi, E., Cicchini, G.~M., Cinelli, L., Biagi, L., Rizzo, S., and Morrone,
  M.~C. (2016).
\newblock Visual {BOLD} {Response} in {Late} {Blind} {Subjects} with {Argus}
  {II} {Retinal} {Prosthesis}.
\newblock {\em PLOS Biology}, 14(10):e1002569.
\newblock Publisher: Public Library of Science.

\bibitem[Chen et~al., 2020]{chen_shape_2020}
Chen, X., Wang, F., Fernandez, E., and Roelfsema, P.~R. (2020).
\newblock Shape perception via a high-channel-count neuroprosthesis in monkey
  visual cortex.
\newblock {\em Science}, 370(6521):1191--1196.
\newblock Publisher: American Association for the Advancement of Science
  Section: Research Article.

\bibitem[Cruz et~al., 2013]{cruz_argus_2013}
Cruz, L.~d., Coley, B.~F., Dorn, J., Merlini, F., Filley, E., Christopher, P.,
  Chen, F.~K., Wuyyuru, V., Sahel, J., Stanga, P., Humayun, M., Greenberg,
  R.~J., Dagnelie, G., and Group, f. t. A. I.~S. (2013).
\newblock The {Argus} {II} epiretinal prosthesis system allows letter and word
  reading and long-term function in patients with profound vision loss.
\newblock {\em British Journal of Ophthalmology}, 97(5):632--636.
\newblock Publisher: BMJ Publishing Group Ltd Section: Clinical science.

\bibitem[da~Cruz et~al., 2018]{da_cruz_phase_2018}
da~Cruz, L., Fynes, K., Georgiadis, O., Kerby, J., Luo, Y.~H., Ahmado, A.,
  Vernon, A., Daniels, J.~T., Nommiste, B., Hasan, S.~M., Gooljar, S.~B., Carr,
  A.-J.~F., Vugler, A., Ramsden, C.~M., Bictash, M., Fenster, M., Steer, J.,
  Harbinson, T., Wilbrey, A., Tufail, A., Feng, G., Whitlock, M., Robson,
  A.~G., Holder, G.~E., Sagoo, M.~S., Loudon, P.~T., Whiting, P., and Coffey,
  P.~J. (2018).
\newblock Phase 1 clinical study of an embryonic stem cell–derived retinal
  pigment epithelium patch in age-related macular degeneration.
\newblock {\em Nature Biotechnology}, 36(4):328--337.

\bibitem[Dagnelie et~al., 2007]{dagnelie_real_2007}
Dagnelie, G., Keane, P., Narla, V., Yang, L., Weiland, J., and Humayun, M.
  (2007).
\newblock Real and virtual mobility performance in simulated prosthetic vision.
\newblock {\em Journal of Neural Engineering}, 4(1):S92.

\bibitem[Dorn et~al., 2013]{dorn_detection_2013}
Dorn, J.~D., Ahuja, A.~K., Caspi, A., da~Cruz, L., Dagnelie, G., Sahel, J.~A.,
  Greenberg, R.~J., McMahon, M.~J., and Grp, A. I.~S. (2013).
\newblock The {Detection} of {Motion} by {Blind} {Subjects} {With} the
  {Epiretinal} 60-{Electrode} ({Argus} {II}) {Retinal} {Prosthesis}.
\newblock {\em Jama Ophthalmology}, 131(2):183--189.

\bibitem[Efron, 1973]{efron_conservation_1973}
Efron, R. (1973).
\newblock Conservation of temporal information by perceptual systems.
\newblock {\em Perception \& Psychophysics}, 14(3):518--530.

\bibitem[Fernandez, 2018]{fernandez_development_2018}
Fernandez, E. (2018).
\newblock Development of visual {Neuroprostheses}: trends and challenges.
\newblock {\em Bioelectronic Medicine}, 4(1):12.

\bibitem[Geruschat et~al., 2015]{geruschat_flora_2015}
Geruschat, D.~R., Flax, M., Tanna, N., Bianchi, M., Fisher, A., Goldschmidt,
  M., Fisher, L., Dagnelie, G., Deremeik, J., Smith, A., Anaflous, F., and
  Dorn, J. (2015).
\newblock {FLORA}™: {Phase} {I} development of a functional vision assessment
  for prosthetic vision users.
\newblock {\em Clinical and Experimental Optometry}, 98(4):342--347.

\bibitem[Granley and Beyeler, 2021]{granley_computational_2021}
Granley, J. and Beyeler, M. (2021).
\newblock A {Computational} {Model} of {Phosphene} {Appearance} for
  {Epiretinal} {Prostheses}.
\newblock In {\em 2021 43rd {Annual} {International} {Conference} of the {IEEE}
  {Engineering} in {Medicine} {Biology} {Society} ({EMBC})}, pages 4477--4481.
\newblock ISSN: 2694-0604.

\bibitem[Hamel, 2006]{hamel_retinitis_2006}
Hamel, C. (2006).
\newblock Retinitis pigmentosa.
\newblock {\em Orphanet Journal of Rare Diseases}, 1(1):40.

\bibitem[Han et~al., 2021]{han_deep_2021}
Han, N., Srivastava, S., Xu, A., Klein, D., and Beyeler, M. (2021).
\newblock Deep {Learning}–{Based} {Scene} {Simplification} for {Bionic}
  {Vision}.
\newblock In {\em Augmented {Humans} {Conference} 2021}, {AHs}'21, pages
  45--54, New York, NY, USA. Association for Computing Machinery.

\bibitem[Hayes et~al., 2003]{hayes_visually_2003}
Hayes, J.~S., Yin, V.~T., Piyathaisere, D., Weiland, J.~D., Humayun, M.~S., and
  Dagnelie, G. (2003).
\newblock Visually guided performance of simple tasks using simulated
  prosthetic vision.
\newblock {\em Artif Organs}, 27(11):1016--28.

\bibitem[Horsager et~al., 2010]{horsager_spatiotemporal_2010}
Horsager, A., Greenberg, R.~J., and Fine, I. (2010).
\newblock Spatiotemporal {Interactions} in {Retinal} {Prosthesis} {Subjects}.
\newblock {\em Investigative Ophthalmology \& Visual Science},
  51(2):1223--1233.

\bibitem[Horsager et~al., 2009]{horsager_predicting_2009}
Horsager, A., Greenwald, S.~H., Weiland, J.~D., Humayun, M.~S., Greenberg,
  R.~J., McMahon, M.~J., Boynton, G.~M., and Fine, I. (2009).
\newblock Predicting {Visual} {Sensitivity} in {Retinal} {Prosthesis}
  {Patients}.
\newblock {\em Investigative Ophthalmology \& Visual Science},
  50(4):1483--1491.

\bibitem[Hou et~al., 2024a]{hou_axonal_2024}
Hou, Y., Nanduri, D., Granley, J., Weiland, J.~D., and Beyeler, M. (2024a).
\newblock Axonal stimulation affects the linear summation of single-point
  perception in three {Argus} {II} users.
\newblock {\em Journal of Neural Engineering}, 21(2):026031.
\newblock Publisher: IOP Publishing.

\bibitem[Hou et~al., 2024b]{hou_predicting_2024}
Hou, Y., Pullela, L., Su, J., Aluru, S., Sista, S., Lu, X., and Beyeler, M.
  (2024b).
\newblock Predicting the {Temporal} {Dynamics} of {Prosthetic} {Vision}.
\newblock In {\em 2024 46th {Annual} {International} {Conference} of the {IEEE}
  {Engineering} in {Medicine} and {Biology} {Society} ({EMBC})}, pages 1--4.
\newblock ISSN: 2694-0604.

\bibitem[Jung et~al., 2024]{jung_stable_2024}
Jung, T., Zeng, N., Fabbri, J.~D., Eichler, G., Li, Z., Willeke, K., Wingel,
  K.~E., Dubey, A., Huq, R., Sharma, M., Hu, Y., Ramakrishnan, G., Tien, K.,
  Mantovani, P., Parihar, A., Yin, H., Oswalt, D., Misdorp, A., Uguz, I.,
  Shinn, T., Rodriguez, G.~J., Nealley, C., Gonzales, I., Roukes, M., Knecht,
  J., Yoshor, D., Canoll, P., Spinazzi, E., Carloni, L.~P., Pesaran, B., Patel,
  S., Youngerman, B., Cotton, R.~J., Tolias, A., and Shepard, K.~L. (2024).
\newblock Stable, chronic in-vivo recordings from a fully wireless
  subdural-contained 65,536-electrode brain-computer interface device.
\newblock Pages: 2024.05.17.594333 Section: New Results.

\bibitem[Kasowski and Beyeler, 2022]{kasowski_immersive_2022}
Kasowski, J. and Beyeler, M. (2022).
\newblock Immersive {Virtual} {Reality} {Simulations} of {Bionic} {Vision}.
\newblock In {\em Augmented {Humans} 2022}, pages 82--93, Kashiwa, Chiba Japan.
  ACM.

\bibitem[Kasowski et~al., 2023]{kasowski_systematic_2023}
Kasowski, J., Johnson, B.~A., Neydavood, R., Akkaraju, A., and Beyeler, M.
  (2023).
\newblock A systematic review of extended reality ({XR}) for understanding and
  augmenting vision loss.
\newblock {\em Journal of Vision}, 23(5):5.

\bibitem[Lorach et~al., 2015]{lorach_photovoltaic_2015}
Lorach, H., Goetz, G., Smith, R., Lei, X., Mandel, Y., Kamins, T., Mathieson,
  K., Huie, P., Harris, J., Sher, A., and Palanker, D. (2015).
\newblock Photovoltaic restoration of sight with high visual acuity.
\newblock {\em Nat Med}, 21(5):476--82.

\bibitem[Luo and da~Cruz, 2016]{luo_argusr_2016}
Luo, Y.~H. and da~Cruz, L. (2016).
\newblock The {Argus}(({R})) {II} {Retinal} {Prosthesis} {System}.
\newblock {\em Prog Retin Eye Res}, 50:89--107.

\bibitem[Mowad et~al., 2020]{mowad_compensatory_2020}
Mowad, T.~G., Willett, A.~E., Mahmoudian, M., Lipin, M., Heinecke, A., Maguire,
  A.~M., Bennett, J., and Ashtari, M. (2020).
\newblock Compensatory {Cross}-{Modal} {Plasticity} {Persists} {After} {Sight}
  {Restoration}.
\newblock {\em Frontiers in Neuroscience}, 14.
\newblock Publisher: Frontiers.

\bibitem[Musk and Neuralink, 2019]{musk_integrated_2019}
Musk, E. and Neuralink (2019).
\newblock An {Integrated} {Brain}-{Machine} {Interface} {Platform} {With}
  {Thousands} of {Channels}.
\newblock {\em Journal of Medical Internet Research}, 21(10):e16194.
\newblock Company: Journal of Medical Internet Research Distributor: Journal of
  Medical Internet Research Institution: Journal of Medical Internet Research
  Label: Journal of Medical Internet Research Publisher: JMIR Publications
  Inc., Toronto, Canada.

\bibitem[Nadolskis et~al., 2024]{nadolskis_aligning_2024}
Nadolskis, L., Turkstra, L.~M., Larnyo, E., and Beyeler, M. (2024).
\newblock Aligning {Visual} {Prosthetic} {Development} {With} {Implantee}
  {Needs}.
\newblock {\em Translational Vision Science \& Technology}, 13(11):28.

\bibitem[Palanker et~al., 2020]{palanker_photovoltaic_2020}
Palanker, D., Le~Mer, Y., Mohand-Said, S., Muqit, M., and Sahel, J.~A. (2020).
\newblock Photovoltaic {Restoration} of {Central} {Vision} in {Atrophic}
  {Age}-{Related} {Macular} {Degeneration}.
\newblock {\em Ophthalmology}.

\bibitem[Paraskevoudi and Pezaris, 2019]{paraskevoudi_eye_2019}
Paraskevoudi, N. and Pezaris, J.~S. (2019).
\newblock Eye {Movement} {Compensation} and {Spatial} {Updating} in {Visual}
  {Prosthetics}: {Mechanisms}, {Limitations} and {Future} {Directions}.
\newblock {\em Frontiers in Systems Neuroscience}, 12.

\bibitem[Parikh et~al., 2010]{parikh_saliency-based_2010}
Parikh, N., Itti, L., and Weiland, J. (2010).
\newblock Saliency-based image processing for retinal prostheses.
\newblock {\em Journal of Neural Engineering}, 7(1):016006.
\newblock Publisher: IOP Publishing.

\bibitem[Prem~Senthil et~al., 2017]{prem_senthil_seeing_2017}
Prem~Senthil, M., Khadka, J., and Pesudovs, K. (2017).
\newblock Seeing through their eyes: lived experiences of people with retinitis
  pigmentosa.
\newblock {\em Eye}, 31(5):741--748.
\newblock Publisher: Nature Publishing Group.

\bibitem[Pérez~Fornos et~al., 2012]{perez_fornos_temporal_2012}
Pérez~Fornos, A., Sommerhalder, J., da~Cruz, L., Sahel, J.~A., Mohand-Said,
  S., Hafezi, F., and Pelizzone, M. (2012).
\newblock Temporal {Properties} of {Visual} {Perception} on {Electrical}
  {Stimulation} of the {Retina}.
\newblock {\em Investigative Ophthalmology \& Visual Science},
  53(6):2720--2731.

\bibitem[Rashbass, 1970]{rashbass_visibility_1970}
Rashbass, C. (1970).
\newblock The visibility of transient changes of luminance.
\newblock {\em The Journal of Physiology}, 210(1):165--186.

\bibitem[Rasla and Beyeler, 2022]{rasla_relative_2022}
Rasla, A. and Beyeler, M. (2022).
\newblock The {Relative} {Importance} of {Depth} {Cues} and {Semantic} {Edges}
  for {Indoor} {Mobility} {Using} {Simulated} {Prosthetic} {Vision} in
  {Immersive} {Virtual} {Reality}.
\newblock In {\em Proceedings of the 28th {ACM} {Symposium} on {Virtual}
  {Reality} {Software} and {Technology}}, {VRST} '22, pages 1--11, New York,
  NY, USA. Association for Computing Machinery.

\bibitem[Rizzo et~al., 2003]{rizzo_perceptual_2003}
Rizzo, III, J.~F., Wyatt, J., Loewenstein, J., Kelly, S., and Shire, D. (2003).
\newblock Perceptual {Efficacy} of {Electrical} {Stimulation} of {Human}
  {Retina} with a {Microelectrode} {Array} during {Short}-{Term} {Surgical}
  {Trials}.
\newblock {\em Investigative Ophthalmology \& Visual Science},
  44(12):5362--5369.

\bibitem[Russell et~al., 2017]{russell_efficacy_2017}
Russell, S., Bennett, J., Wellman, J.~A., Chung, D.~C., Yu, Z.-F., Tillman, A.,
  Wittes, J., Pappas, J., Elci, O., McCague, S., Cross, D., Marshall, K.~A.,
  Walshire, J., Kehoe, T.~L., Reichert, H., Davis, M., Raffini, L., George,
  L.~A., Hudson, F.~P., Dingfield, L., Zhu, X., Haller, J.~A., Sohn, E.~H.,
  Mahajan, V.~B., Pfeifer, W., Weckmann, M., Johnson, C., Gewaily, D., Drack,
  A., Stone, E., Wachtel, K., Simonelli, F., Leroy, B.~P., Wright, J.~F., High,
  K.~A., and Maguire, A.~M. (2017).
\newblock Efficacy and safety of voretigene neparvovec ({AAV2}-{hRPE65v2}) in
  patients with {RPE65}-mediated inherited retinal dystrophy: a randomised,
  controlled, open-label, phase 3 trial.
\newblock {\em The Lancet}, 390(10097):849--860.
\newblock Publisher: Elsevier.

\bibitem[Sadato et~al., 1996]{sadato_activation_1996}
Sadato, N., Pascual-Leone, A., Grafman, J., Ibañez, V., Deiber, M.-P., Dold,
  G., and Hallett, M. (1996).
\newblock Activation of the primary visual cortex by {Braille} reading in blind
  subjects.
\newblock {\em Nature}, 380(6574):526--528.
\newblock Publisher: Nature Publishing Group.

\bibitem[Sainohira et~al., 2018]{sainohira_quantitative_2018}
Sainohira, M., Yamashita, T., Terasaki, H., Sonoda, S., Miyata, K., Murakami,
  Y., Ikeda, Y., Morimoto, T., Endo, T., Fujikado, T., Kamo, J., and Sakamoto,
  T. (2018).
\newblock Quantitative analyses of factors related to anxiety and depression in
  patients with retinitis pigmentosa.
\newblock {\em PloS One}, 13(4):e0195983.

\bibitem[Sanchez-Garcia et~al., 2020]{sanchez-garcia_semantic_2020}
Sanchez-Garcia, M., Martinez-Cantin, R., and Guerrero, J.~J. (2020).
\newblock Semantic and structural image segmentation for prosthetic vision.
\newblock {\em PLOS ONE}, 15(1):e0227677.

\bibitem[{Second Sight}, 2013]{second_sight_argus_2013}
{Second Sight} (2013).
\newblock {\em Argus® {II} {Retinal} {Prosthesis} {System} {Surgeon}
  {Manual}}.
\newblock Number 900029-001 Rev C. Second Sight Medical Products, Inc., Sylmar,
  CA.

\bibitem[Vergnieux et~al., 2017]{vergnieux_simplification_2017}
Vergnieux, V., Macé, M. J.-M., and Jouffrais, C. (2017).
\newblock Simplification of {Visual} {Rendering} in {Simulated} {Prosthetic}
  {Vision} {Facilitates} {Navigation}.
\newblock {\em Artificial Organs}, 41(9):852--861.
\newblock Publisher: John Wiley \& Sons, Ltd.

\bibitem[Weiland et~al., 2016]{weiland_electrical_2016}
Weiland, J.~D., Walston, S.~T., and Humayun, M.~S. (2016).
\newblock Electrical {Stimulation} of the {Retina} to {Produce} {Artificial}
  {Vision}.
\newblock {\em Annual Review of Vision Science}, 2(1):273--294.

\bibitem[Wilke et~al., 2011]{wilke_electric_2011}
Wilke, R. G.~H., Moghadam, G.~K., Lovell, N.~H., Suaning, G.~J., and Dokos, S.
  (2011).
\newblock Electric crosstalk impairs spatial resolution of multi-electrode
  arrays in retinal implants*.
\newblock {\em Journal of Neural Engineering}, 8(4):046016.

\bibitem[Yücel et~al., 2022]{yucel_factors_2022}
Yücel, E.~I., Sadeghi, R., Kartha, A., Montezuma, S.~R., Dagnelie, G., Rokem,
  A., Boynton, G.~M., Fine, I., and Beyeler, M. (2022).
\newblock Factors affecting two-point discrimination in {Argus} {II} patients.
\newblock {\em Frontiers in Neuroscience}, 16:901337.

\end{thebibliography}

\clearpage
\appendix

\section{Eye-Tracking Accuracy of the HTC Vive Pro}
\label{sec:app-eye-tracking}

To assess the precision of the HTC Vive Pro's built-in eye tracker, $n=30$ participants tracked a moving on-screen dot ($\sim 2.4^\circ$ visual angle) using their eyes. 
The dot moved randomly between four corners positioned halfway between the center and edges of the screen. 
It traversed the distance between points over $2.5 \pm 0.5$ seconds and remained stationary at each location for 1.5 seconds.

The angular error, defined as the distance between the dot's center and the user's gaze location, was measured every \SI{0.1}{\second}. 
Measurements were taken during both fixation (when the dot was stationary) and pursuit (when it was moving). 
Mean angular error was \SI{1.904 \pm 2.048}{\degree} during fixation and \SI{1.838 \pm 1.660}{\degree} during pursuit, with no significant difference between the two conditions (t-test for non-equal variances, $p > 0.27$).

\begin{figure}[h!]
    \centering
    \includegraphics[width=\linewidth]{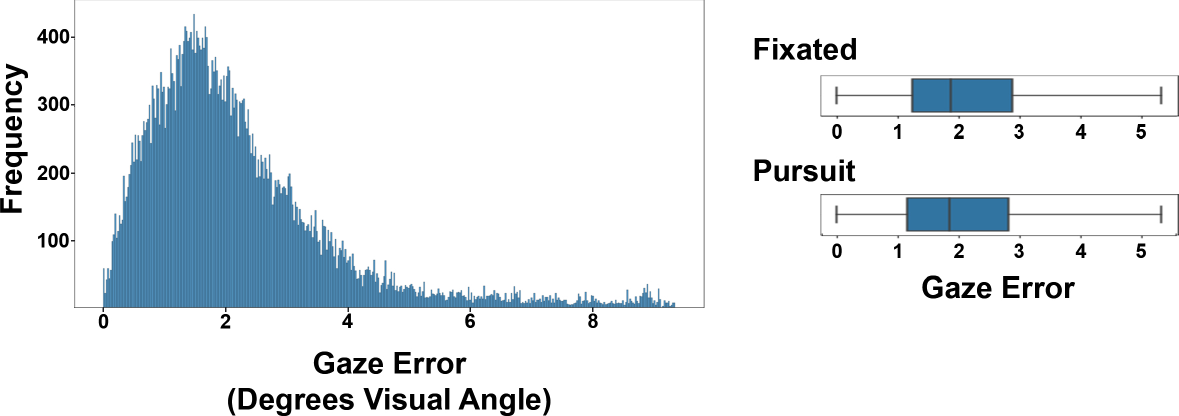}
    \caption{
        \textbf{Eye tracking accuracy of the HTC Vive Pro.} 
        The histogram (left) shows the distribution of angular gaze error (degrees visual angle) across all measurements, with most errors falling below \SI{5}{\degree}. 
        The boxplots (right) compare gaze error during fixation (when the dot was stationary) and pursuit (when the dot was moving). 
        Mean errors were similar across conditions (\SI{1.904 \pm 2.048}{\degree} for fixation and \SI{1.838 \pm 1.660}{\degree} for pursuit), with no significant difference between the two (t-test, $p > 0.27$). 
        Over 94\% of measurements had an error below \SI{5}{\degree}, confirming the system's precision for gaze-contingent rendering.
    }
    \label{fig:eye-tracking}
\end{figure}

Overall, 94.1\% of measurements had an angular error below \SI{5}{\degree}, and 80\% were below \SI{3}{\degree}. 
These results indicate that the HTC Vive Pro provides adequate precision for gaze-contingent rendering in simulated prosthetic vision experiments (Figure~\ref{fig:eye-tracking}).

\end{document}